\documentclass[aps,prl,preprint,superscriptaddress,floatfix, %linenumbers
]{revtex4-2}
\usepackage[utf8]{inputenc}
\usepackage{amsmath}
\usepackage{verbatim}
\usepackage{natbib}
\usepackage{graphicx}
\usepackage{xcolor}

\setcitestyle{super}

\usepackage{booktabs}

\usepackage{multirow}
\usepackage{xcolor}

\begin{document}
\title{Excitonic Resonances in Coherent Anti-Stokes Raman Scattering from Single Wall Carbon Nanotubes}

\author{Georgy Gordeev}
\email{georgy.gordeev@uni.lu}
\affiliation{Departamento de Física, Universidade Federal de Minas Gerais, Belo Horizonte, Minas Gerais 30123-970, Brazil}
\affiliation{Freie Universit\"at Berlin, Department of Physics, Arnimallee 14, 14195 Berlin}
\affiliation{Department of Physics and Materials Science, University of Luxembourg, L-4422 Belvaux, Luxembourg}

\author{Lucas Lafeta}
\affiliation{Departamento de Física, Universidade Federal de Minas Gerais, Belo Horizonte, Minas Gerais 30123-970, Brazil}
\affiliation{Ludwig-Maximilians-Universität München, Department of Chemistry and Center for NanoScience (CeNS), Butenandtstraße 5-13 (E), 81377 Munich, Germany}

\author{Benjamin S. Flavel}
\affiliation{Institute of Nanotechnology, Karlsruhe Institute of Technology, 76201 Eggenstein-Leopoldshafen, Germany}
\author{Ado Jorio}
\affiliation{Departamento de Física, Universidade Federal de Minas Gerais, Belo Horizonte, Minas Gerais 30123-970, Brazil}
\author{Leandro M. Malard}
\email{lmalard@fisica.ufmg.br}
\affiliation{Departamento de Física, Universidade Federal de Minas Gerais, Belo Horizonte, Minas Gerais 30123-970, Brazil}

\begin{abstract}
    In this work, we investigate the role of exciton resonances in coherent anti-Stokes Raman scattering (er-CARS) in single-walled carbon nanotubes (SWCNTs). We drive the nanotube system in simultaneous phonon and excitonic resonances and we observe a superior enhancement by orders of magnitude exceeding non-resonant cases. We investigated the resonant effects in several $(n,m)$ chiralities and found that the er-CARS intensity varies more than four orders of magnitude between nanotube species determined by excitonic resonant condition. The experimental finding is compared with a perturbation theory model. Finally, we show that such giant resonant non-linear signals enable rapid mapping and local heating of individualized CNTs, suggesting easy tracking of CNTs for future nanotoxicology studies and therapeutic applications in biological tissues.
    
  \end{abstract}
\maketitle

\begin{figure}
    \centering
    \includegraphics[width=9cm]{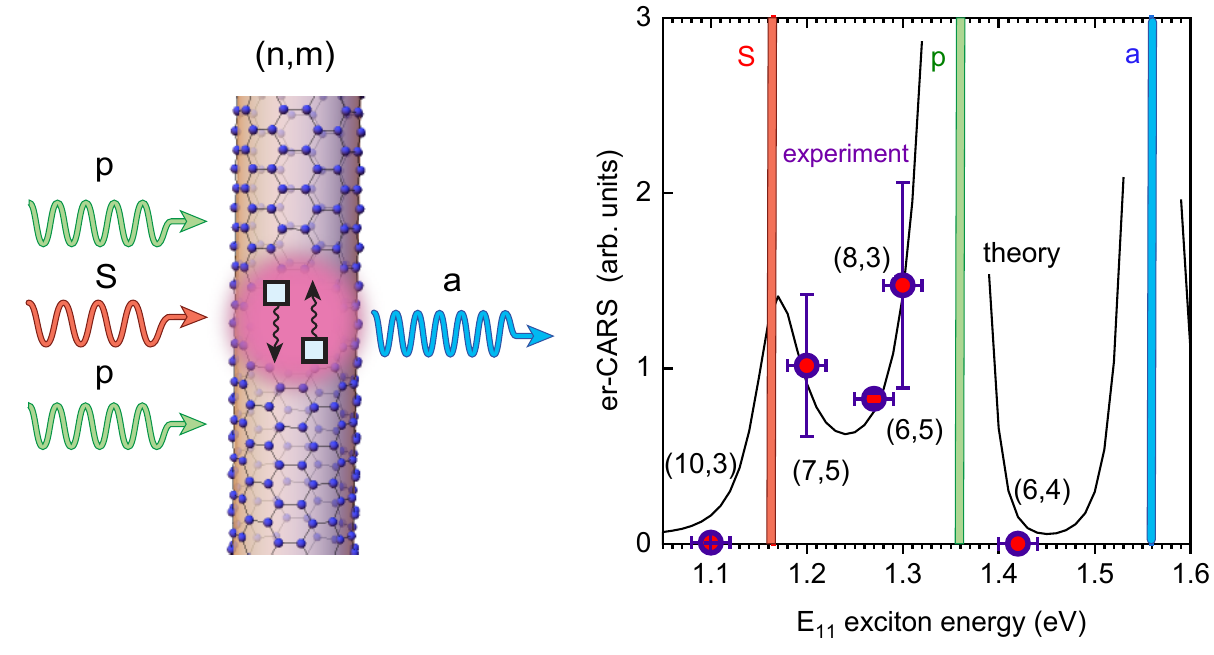}
    %\caption{Caption}
    %\label{fig:my_label}
\end{figure}
\newpage
\section{Introduction}
Raman spectroscopy has been extensively used to characterize materials and biological specimens. However, the Raman effect is intrinsically weak due to the small scattering cross-sections of the materials under investigation\cite{boyd,livroxie,potma2017, malard21, biosrs}. Such low signals prevent the application of the technique for rapid imaging of biological cells, tissues, and in-vivo applications \cite{livroxie,biosrs,Huth2018}. 
On the other hand, non-linear optical effects that resonate with vibrational states have potential to overcome these obstacles, such as Coherent Anti-Stokes Raman Spectroscopy (CARS) and Stimulated Raman Scattering. The application of such techniques in combination with nanomaterials may advance drug delivery, object tracking, and tumor therapy\cite{potma2017,malard21,nanotubebio1} when additional enhancement through excitonic resonance can be achieved. In this sense, single-walled carbon nanotubes (SWCNTs) are good canditates as they can easily penetrate living cells due to their unique aspect ratio \cite{ntcell} and have pronounced excitonic properties\cite{Wang2005}. Here we show that by simultaneously driving the SWCNTs into resonance with exciton and phonon states, the CARS intensity increases exceeding non-resonant case by orders of magnitude. Moreover, we studied different nanotube chiralities and found that the CARS enhancement is sensitive to the energy of the excitonic resonance, which can be modeled by a perturbation theory. In order to explore the applicability of CNTs for tracking or future nanotoxicology studies \cite{nanotoxology} we scan individual tubes deposited with up to 16 $\mu s$ per pixel imaging rates and heat them locally by elevated laser powers up to 200 °C.\par
%, whereas the use of nanomaterials in tracking and therapy is highly desirable
%Therapeutic applications \cite{}

Coherent anti-Stokes Raman scattering can be interpreted as a four-wave mixing (FWM) process resonating with a phonon. It combines four photons, two of equal energy (pump) and one lower energy (Stokes) incident in the sample and the fourth photon emerging from it. In CARS, the energy difference between the Pump and Stokes beams is tuned to match a specific vibrational mode, enhancing the FWM signal by the anti-Stokes process  \cite{boyd,livroxie}. Furthermore, if the energy of one of the beams matches real excitonic or electronic transition, the CARS process becomes electronically (excitonically) resonant, which we refer to as excitonic resonant CARS (er-CARS)\cite{ecars1}. This method has been successfully applied to molecules, significantly improving the detection sensitivity of the induced Raman signal \cite{esrs1}. Due to quantum confinement, nanomaterials can have narrow bandwidth excitonic resonances and thus are promising candidates for resonant er-CARS effects. However, such an application has so far been limited. For example, CARS has been used to probe nanomaterials inside biological specimens but without the use of resonant effects. In the case of 2D materials, the broad electronic contribution of graphene leads to the interference between the FWM and CARS signals, making it hard to use such technique for imaging proposes \cite{Lafeta2017,virga2019coherent,malard21}. On the other hand, the optical properties of semiconducting SWCNTs are dictated by the excitonic properties only, with single particle contributions being negligible. 

Single-walled carbon nanotubes are tubular structures, where one-dimensional confinement yields excitonic states with high energy\cite{Luer2009} phonons of $\sim0.19$ eV with large exciton-phonon coupling.\cite{Doorn2008,Jiang2007a} For instance, this reflects in high resonance Raman cross sections,\cite{Maruyama2010,Tschannen2020} where single tubes can be experimentally measured by Raman spectroscopy \cite{Jorio2002,Cronin2005}. The linear absorption of CNTs is dominated by sharp excitonic resonances at room temperature, which are determined by the nanotube chiral indices.\cite{Weisman2003}. CARS has been used to study carbon nanotubes, either using  raw soot ensembles of different chiralities \cite{Paddubskaya:18,Duarte2013} or individual tubes grown by chemical vapor deposition\cite{Sheps2012,Kim2009}.  In both cases, the excitonic resonant states were not well defined since the CNT were not a single chirality, and therefore their electronic states overlap. Recently the chirality separation techniques became available that can enrich a pure single chirality \cite{Liu2011,Flavel2014,Flavel2013}. Here we show that such samples allow the use of er-CARS in resonance with excitons in SWCNT.

\section{Experimental Methods}

%Due to the high affinity of (6,5) to the Sephacryl-S200 gel (Amersham Biosciences) and their ability to displace other (n,m) species at 1.6 wt \%  SDS (Merck), 1 wt \% sodium cholate (SC, $\gtrsim$ 99 \%, Sigma Aldrich) was used as an eluent in a one-column approach without the use of a pH gradient.
The nanotube \textit{chirality-enriched samples} of either (6,5), (8,3), (6,4), (7,5)\cite{Flavel2014} or (10,3)\cite{ Liu2011} were prepared by the gel permeation method from the HiPco raw material (NanoIntegris). The (6,5) SWCNTs were crafted from HiPco raw material (NanoIntegris) as outlined previously using a gel permeation chromatography system.\cite{Flavel2014,Flavel2013}  The nanotube suspension was drop cased onto the thin quartz substrate. The dried samples were put in a water beaker overnight to remove surfactant residues. For \textit{resonant CARS measurements}, a picosecond laser (APE PicoEMERALD) was used. This laser system generated two laser beams that can be spatially and temporally overlapped with one wavelength fixed at 1064 nm $(E_{S})$ (1.17 eV) with 7 ps pulse duration and the second beening tunable between 730 nm (1.7 eV) and 960 nm (1.27 eV) ($E_{p}$) with 5–6 ps pulse duration. The $E_{p}$ is generated by pumping the optical parametric oscillator cavity with the second harmonic of the 1064 nm laser at 532 nm. The selected $E_{p}$, $E_{S}$ lasers were directed to the sample by mirrors. A beam-splitter separated the collection and detection parts. Afterward, the anti-Stokes signal $E_{a}$ was filtered by a short-pass filter and focused on a slit of the (Andor Shamrock SR-303i-B) spectrometer equipped with a charge-coupled device (iDus DU420A-BEX2-DD). The laser powers on the sample were kept below $\sim$0.5 mW for resonant experiments shown in Figures \ref{FIG:Mtd}, \ref{FIG:FWMnm}. Only for the measurement shown in Figure \ref{FIG:Appl}c we used a higher laser power, as indicated in the Figure.

%\begin{comment}
%and between $\sim$1193 and $\sim$1961 nm $(\omega{}Idler)$,
%\end{comment}
The same detection setup was used for \textit{Raman measurements}. The diode laser at 561 nm (2.2 eV) was used for sample excitation, combined with a long-pass filter to separate Raman from Rayleigh scattering. For simultaneous measurements with Raman and CARS, the visible and IR beams were aligned using a beam splitter and focused on the same spot on a sample. 

In order to \textit{calibrate er-CARS intensity}, we needed to account first for the difference in laser power used, and second for the different concentration of CNTs under the laser spot. The CARS signals are proportional to $P_{tot} =I_{p}^2I_{S}$. Therefore the integrated intensity of the CARS was first divided by $P_{tot}$. We utilize Raman spectroscopy to determine the concentration of the CNTs under the laser spot. When the laser energy matches the transition energy, the G mode Raman intensity of one individual CNT is predicted to weakly depend on chirality and diameter due to the similarity in matrix elements.\cite{Jiang2007a} We account for the different resonance energies between (n,m) by introducing an intensity factor $f_{Ram}$, which depends on the energy difference between $E_{22}$ and $E_{las}$. This dependence is chirality specific for the G mode however it is known from previous experiments for all the species except for the (10,3) SWCNTs, which we approximate to the (8,3) SWCNTs.\cite{Duque2011,Gordeev2017} The CNTs concentration $C$ is then $I_{Ram}f_{Ram}$. Finally, the Raman intensity scales with $\sim C$, but CARS intensity is proportional to the $C^2$. To show resonant effects, we then calibrate $I_{CARS}$ as $I_{CARS}^{measured}/P_{tot}/C^2$, accounting for laser powers and different concentrations. The error bars in Figure \ref{FIG:Theory} were computed using standard error propagation methods. These mainly originate from uncertainty in estimating the $E_{22}$ transition energy influencing the resonant factor of Raman scattering.

For the CARS intensity image, we have redirected the $E_{S}$ and $E_{p}$ pulses to a scanning laser microscope (LaVision Biotec). The scanning mirrors allow the beams to map the sample when focused by the 60× apochromatic air objective (numerical aperture of 0.95) in the scanning laser microscope. The backscattered CARS signal is then directed to the photomultiplier. A short pass filter eliminates laser contamination. The average scan contained $10^6$ pixels and was completed within 16s.

\section{Results}

%Fig1%

\begin{figure}
  \centering
  \includegraphics[width=8cm]{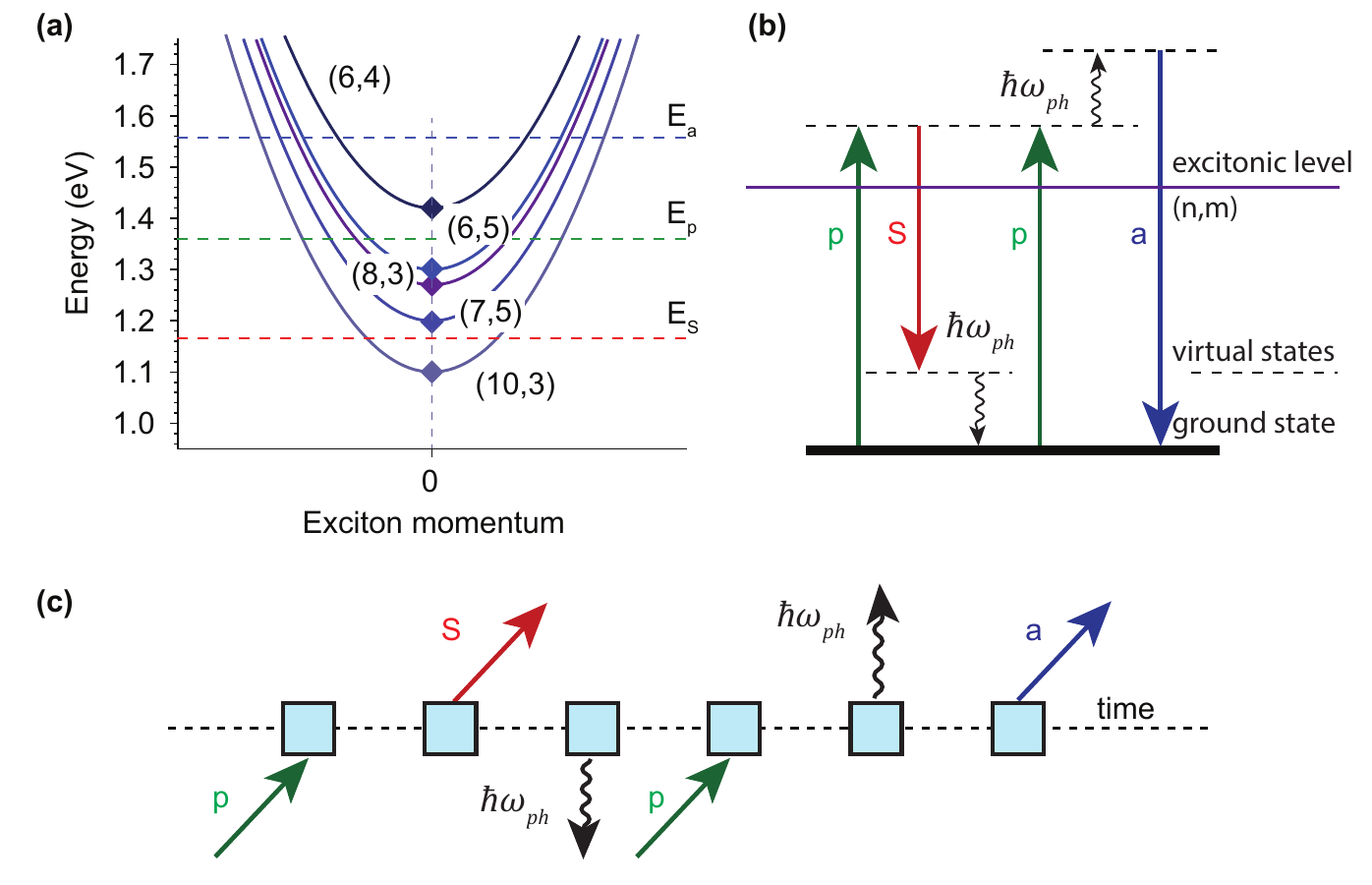}
  \caption{ The concept of the er-CARS scattering in SWCNTs. (a) Exciton dispersion for different (n,m) SWNT probed in this study, along with fixed energies of the pump ($E_p$), Stokes ($E_S$), and anti-Stokes beams ($E_a$). (b) Scheme of one er-CARS scattering pathway in an excitonic system. (c) Feynman diagram representing sixth-order process, straight (curly) arrows indicate an interaction with a photon (phonon); see discussion in text for possible permutations. %\textcolor{red}{Check the order of interaction in (c). Stokes emission should be third and phonon creation in second.}
  }
  \label{FIG:sch}
\end{figure}
Four-wave mixing comprises the interaction of four photons; incident three photons can excite or de-excite the system yielding an emission of the fourth photon. In CARS two incident have the same energy $E_p$ and originate from the pump beam. The third Stokes photon $E_S$ has a smaller energy. The energy of the fourth anti-Stokes photon is pre-determined by the first three and yields $E_a=2E_p-E_S$ \cite{boyd}. In our setup the energy of the Stokes beam is kept constant $E_S=$1.165 eV, whereas the energy of the pump beam can be adjusted to match particular phonon. The total non-linear enhancement has vibrational and excitonic components 
\begin{equation}
\chi^{(3)}_{total}= \chi^{(3)}_{er-CARS} + \chi^{(3)}_{vib},% \left[\sum_{k}^{43}{I_k} \right]^2,
\label{EQ:chi_total}
\end{equation}
with signal scaling as $ |\chi^{(3)}_{total}|^2$. The $\chi^{(3)}_{er-CARS}$ depends on the detuning from the exciton, and the vibrational depends on phonon detuning $\Delta_{vib} = E_p-E_S$ as \cite{Hudson1976} 
\begin{equation}
\chi^{(3)}_{vib}=\sum_{ph}{\frac{\chi_{ph}}{\hbar\omega_{ph}-\Delta_{vib} -i\Gamma_{ph}} },
\label{EQ:vib}
\end{equation}
where $\Gamma_{ph}$ is a phonon lifetime broadening. $\chi_{ph}$ is electric dipole transition moment of the phonon mode. Different vibrational states can be resonantly excited by matching the energy $E_p-E_S$ with phonon energy $E_{ph}$.\cite{livroxie}\par

The excitonic enhancement factor $\chi^{(3)}_{er-CARS}$ in a one-dimensional excitonic system requires detailed treatment. Intuitively, one would expect a more efficient CARS process whenever the $E_u, (u=p,S,a)$  meets the excitonic resonance of the nanotube (er-CARS). The photon energies for $E_p$, $E_S$, and $E_a$ are compared with the exciton energies used in Figure \ref{FIG:sch}a, where it is possible to observe different resonances. We therefore needed a model to describe the influence of resonant states explicitly \cite{boyd}. Historically, four-wave mixing was mostly studied in zero-dimensional systems, such as molecules. In such systems, the excitonic and vibrational states are equal in treatment, and a density matrix approach is typically used to describe non-linear interactions with light.\cite{mukamel1999} However, in solid state systems, the excitonic and phonon states are independent quasi-particles, making this approach not applicable\cite{Yu1995}. Instead, the interaction between phonons states and electronic excitation occurs via electron-phonon Hamiltonian, where the electron (exciton) can be scattered to a new state by creating or emitting a phonon. Similarly, excitons can be scattered to new states by photon absorption or emission through the exciton-photon Hamiltonian. To understand the interplay between different matrix elements and scattering pathways, we turn to perturbation theory, widely used to describe resonant Raman processes in solids.\cite{Yu1995} 

Figure \ref{FIG:sch}b depicts one possible CARS scattering process, where the energy difference between pump and Stokes equals optical phonon energy $\hbar \omega_{ph}$. The dashed horizontal lines represent the virtual states of the entire system. These virtual states can resonate with the excitonic level of different SWCNTs that depend on chiral indices (n,m). Unlike the molecular CARS diagram, we draw additional phonon emission and absorption steps. The same process is shown in Figure \ref{FIG:sch}c as a Feynman diagram. Each vertex of the Feynman diagram represents an interaction with a photon or phonon; the vertices are ordered in time. An arrow pointing to the system represents an energy increase, whereas an arrow pointing away from the system represents an energy decrease. The phonon creation and eliminations occur at the vertices two and five, respectively. By walking through the steps first \textit{(i)}, we see that the system is excited by the $E_p$. Later on \textit{(ii)} the interaction with a negative component of $E_S$ electric field de-excites the system towards the ground state, and \textit{(iii)} the phonon is created $E_{ph}$. Next, the third photon $E_p$ excites the system into a higher state \textit{(iv)}, and after it, the phonon that was created before is destroyed \textit{(v)}. Finally, the fourth wave photon emerges from the system \textit{(vi)}, bringing the system back to the ground state, which satisfies the energy conservation principle built into the perturbation theory. \par

%Fig2%
\begin{figure}
  \centering
  \includegraphics[width=16cm]{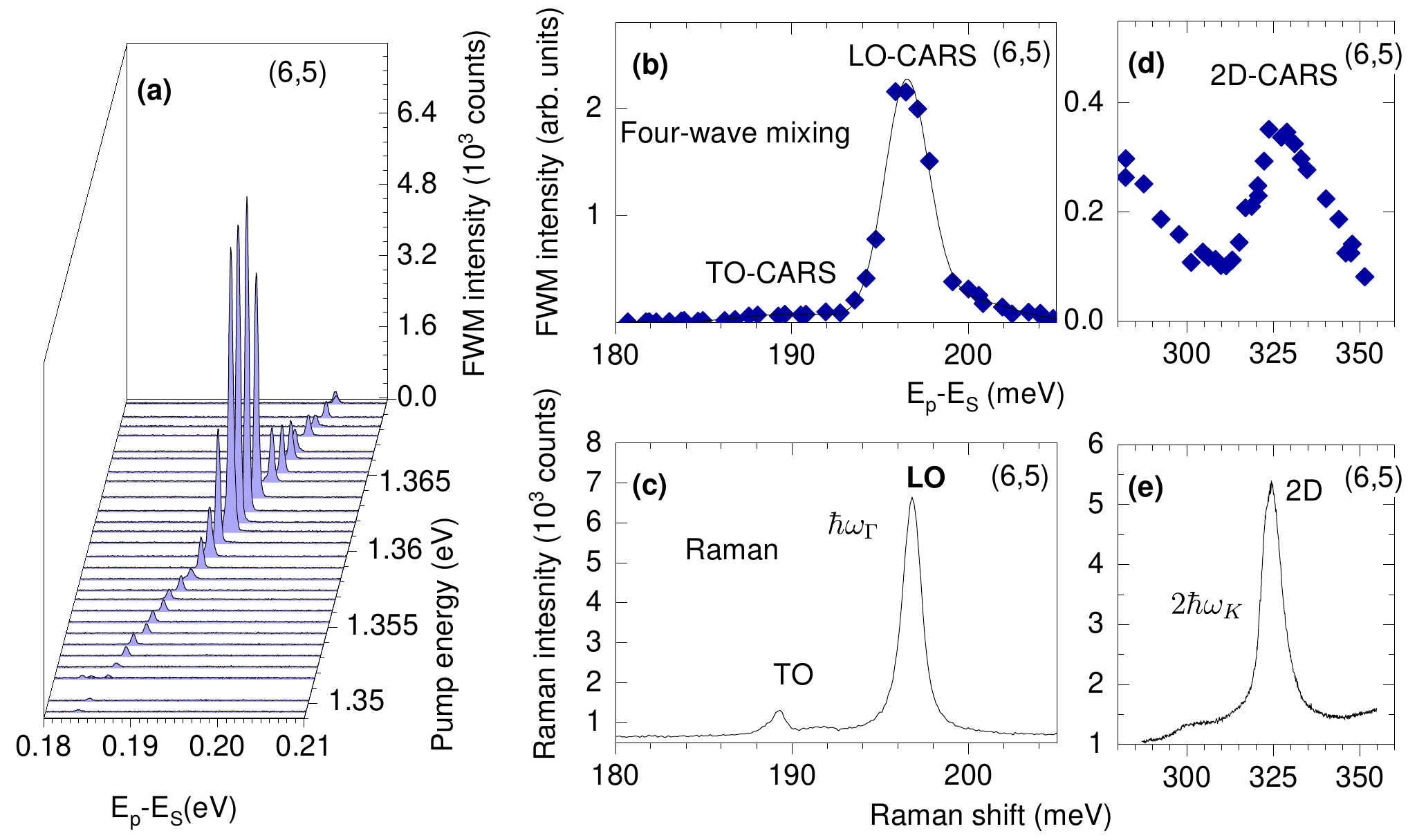}
  \caption{Four-wave mixing (FWM) and Raman scattering in the (6,5) CNT across one- and two-phonon resonances. (a) Waterfall plot of the experimental FWM spectra, the $x$ axis shows the difference between anti-Stokes and pump energies, $y$ axis represents the pump energy $E_p$.  (b)  Integrated area of the $E_a$ signal vs phonon energy. (c) Raman scattering intensity comparison between the longitudinal and transverse resonances. (d) FWM profile of across the 2D mode and (e) Raman spectrum in the same energy range. }
  \label{FIG:Mtd}
\end{figure}

Quantum interference between many scattering processes determines the total er-CARS signal enhancement.  In our calculation, we permute vertices and discard all processes that yield negative energy at any of the steps. For example, phonon absorption cannot be the initial steps of the process and all pathways with such a step are discarded from the sum in Eq. \eqref{EQ:sum}. As a result the entire permutation group consisting of 6!/2! members reduces to 43. We impose the phonon emission and scattering to be equally probable since a large number of photons are incident on the sample during a pico-second laser pulse. Each of the 43 processes contributes to the er-CARS intensity with equal probability yielding an interference effect:

\begin{equation}
\chi^{(3)}_{er-CARS}= \sum_{k}^{43}{I_k} % \left[\sum_{k}^{43}{I_k} \right]^2,
\label{EQ:sum}
\end{equation}
where $I_k$ is the contribution of a $k$-th scattering pathway. The square is taken after summation allowing the processes to interfere. The $I_k$ can be expressed as:

\begin{equation}
\begin{split}
  I_k(E_{11}[n,m])=\frac{M_{ex-q}^2  (M_{ex-ph})^4 }{ \prod_{l=1}^{5}(E_{11}[n,m]+\sum_{j=1}^{l}{\pm\hbar\omega_{j}}-i\gamma_{11})
   }.
  \label{EQ:member}
\end{split}
\end{equation}

\noindent $M_{ex-ph} $ and $M_{ex-q} $ are exciton-photon and exciton-phonon matrix elements, respectively. $\hbar\omega_j$ is the energy quanta interacting with the \textit{j}-th vertex, either photon($E_p,E_S,E_a$) or phonon $E_{ph}$. $\gamma_{11}$ is the broadening factor related to the excitonic lifetime. The sign of $\hbar\omega_i$ is determined by the direction of the arrow, either pointing away or towards the system in a Feynman diagram; see example in Figure \ref{FIG:sch}c. For one possible process, these energies are depicted in Figure \ref{FIG:sch}b, c. From Eq. \eqref{EQ:sum}, we expect that resonant enhancement of the CARS process when either one of the three involved photons is close to the energy of the exciton. We can verify this experimentally by using different nanotube chiralities (n,m), which would enter Eq. \eqref{EQ:member} as $E_{11}[n,m]$. While (n,m) chiralities with $E_{11}$ exactly matching $E_{p}$ will provide a large enhancement, the most interesting interference effects happen in between $E_S$ and $E_a$. We therefore focus on (n,m) chiralities in this range.\par

At this point, we outline the essential differences between er-CARS in molecules and solids. In molecules, the vibrational levels are treated the same way as the electronic states and phonons are not emitted nor absorbed. A typical perturbation theory approach has a reduced order, compared to Eq. \eqref{EQ:sum} and $\chi^{(3)}_{molecule}$ represents a coherent sum of contributions from the vibrational and electronic states that can be resonant with either $E_p, E_S$, or $E_a$, similar to Eq \eqref{EQ:member}.\cite{Hudson1976} Despite similarities, these expressions are not the same. The matrix elements in the numerator can differ from each other, depending on the excited state displacement coordinate. Further, several vibrational states could be at play and increase the number of outgoing resonances. Finally, the molecular er-CARS theory only describes dispersionless electrons and phonons, which is not the case in CNTs. We now turn to the experimental investigation of Eqs. \eqref{EQ:sum}
 and \eqref{EQ:vib}.\par
 
Vibrational enhancement is superior for the longitudinal phonon compared to the transverse one. We investigate the effect of vibrational enhancement in (6,5) nanotubes. Figure \ref{FIG:Mtd}a shows the FWM signal for the (6,5) SWCNT for several pump energies ($E_p$) spanning across vibrational resonances and for a fixed Stokes energy ($E_S=1.164$ eV). When the pump energy approaches 1.362 eV ($E_p-E_S=$ 0.197 eV), which matches well the longitudinal (LO) phonon energy of the SWCNT, the FWM intensity has a gigantic $\approx 2\cdot 10^3$ times enhancement. Such resonance with the LO phonon is shown in Fig. \ref{FIG:Mtd}b, where the integrated area of the FWM signal is plotted concerning $E_p-E_S$ (or $E_a-E_p$). The enhancement reported here is much more significant than a previously reported 2x on-off resonance contrast to \textit{Kim et al}.\cite{Kim2009}. We attribute this to the excitonic resonance effect, discussed in the following paragraphs. We observe a minor enhancement for the transverse (TO) phonon with $E_{TO}$ close to 0.189 eV (see Fig. \ref{FIG:Mtd}b). The CARS intensity resonant with the LO phonon is $\sim$25 times higher than with the TO phonon. On the other hand, in Raman scattering experiments, the LO phonon is only 5 times more intense than the TO phonon, see Fig. \ref{FIG:Mtd}c. Our perturbation theory captures such a difference in the LO/TO intensity ratio well. Stokes Raman scattering is proportional to $(M_{ex-q})$ since the phonon is only emitted once. On the other hand, the CARS scales as $(M_{ex-q})^2$ because the exciton-phonon interaction occurs twice, when phonon is created and eliminated. Therefore the observed difference for the LO/TO enhancement factor fits well with this theoretical interpretation.\par

With er-CARS, we can collect low intensity signals from double resonant Raman scattering processes outside the Brillouin zone center, see Figure \ref{FIG:Mtd}d. Although much higher laser power is necessary, we can visualize the variation of FWM intensity near the two-phonon resonance of the 2D mode for the first time. Such a process would involve eight steps and is relatively inefficient since two more exciton-phonon matrix elements are involved. Compared to the Raman scattering, the FWM line shape of the 2D mode is asymmetric and is shifted to higher energies, see Figure \ref{FIG:Mtd}e. The 2D mode in SWCNTs is not yet well understood, and the scattering pathways involve scattering of the exciton by the K point phonons\cite{Thomsen2000}. They may include polaritonic effects,\cite{Georgy2019th} therefore, we focus on the FWM profiles produced by a single phonon.

\begin{figure}
  \centering
  \includegraphics[width=8cm]{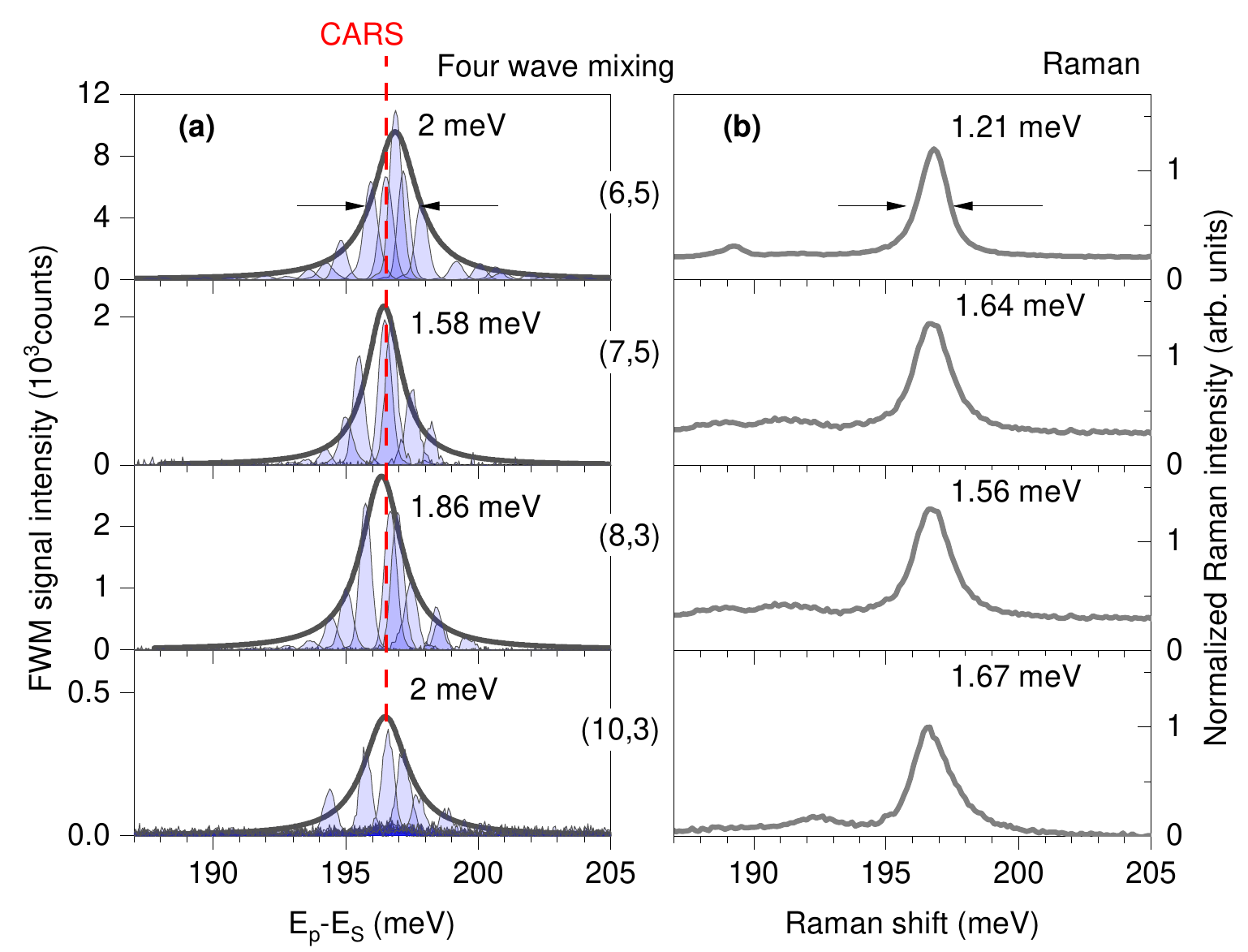}
  \caption{ Four-wave mixing in different (n,m) CNTs in resonance with the LO phonon. (a) FWM profiles in different chiralities, solid lines are \eqref{EQ:vib} fits, and the numbers near the plot give their full width at half maxima. The Stokes and Pump intensities were 60, 150, 100, 590 $\mu W $, from top to bottom, $E_S=1.165$ eV , $E_p=1.35-1.37$ eV. (b) Raman spectra were taken at the same spot, with laser energy of 2.2 eV.}
  \label{FIG:FWMnm}
\end{figure}

The phonon linewidth is slightly different between Raman and four-wave mixing. We measure the FWM signal across the vibrational resonance of the G mode in different nanotube chiralities, with the $E_{11}$ energy spanning between 1.1 eV in (10,3) SWCNTs and 1.42 eV in $(6,4)$ CNTs, as shown in Figure \ref{FIG:Mtd}a. In all the SWCNT chiralities measured, the highest FWM signal was found for resonance with the LO phonon; see red vertical line Figure \ref{FIG:FWMnm}a. The line shape of the FWM profiles is mainly determined by the lifetime of the phonon as $\sim \hbar/\tau_{ph}$. A similar lifetime can be observed in the Raman spectra for the same SWCNT as shown in Figure \ref{FIG:FWMnm}b, where the widths of the Raman G band are compared. Overall, the FWM profiles have higher full widths at half maximum (fwhm) than measured by Raman spectroscopy, except for the (7,5) SWCNTs, where the entire width is almost identical with a value of 1.6 meV. Two factors can explain the difference in line widths. First, we can collect the FWM signal from the larger part of BZ as long as created and annihilated phonons have net zero momentum, and second, we can generate a higher electronic temperature with the pulsed laser excitation as shown by previous works on graphene \cite{Ferrante2018}. In these experiments, the CARS linewidth was broader than Raman spectroscopy,\cite{Lafeta2017,Ferrante2018,virga2019coherent} where the photon flux is moderate and can create hot electron temperatures.

%Fig3%

\begin{figure}
  \centering
  \includegraphics[width=8cm]{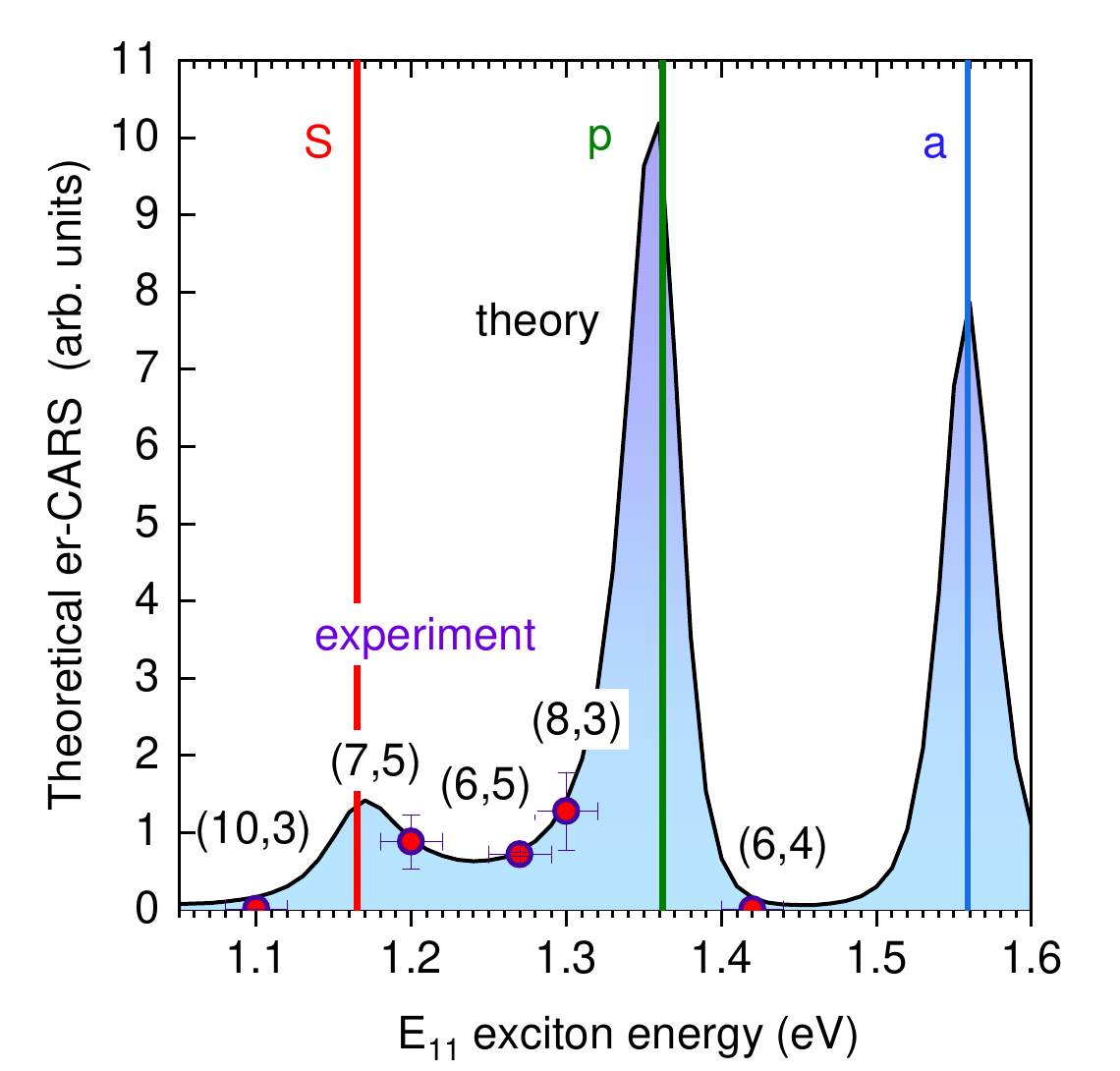}
  \caption{er-CARS intensities for different $(n,m)$. The energies of $E_p$, $E_S$, and $E_{a}$ are fixed whereas the $E_{11}$ energy is varied. Symbols represent experimental data, integrated area of the FWM peak normalized on power, and nanotube concentration obtained through Raman spectroscopy. The line represents theoretical calculation by Eq. \eqref{EQ:sum} with $\gamma_{11}=$ 40 meV, $E_S=$1.165 eV, $E_p=$1.36 eV, and $E_{ph}=$0.196 eV.}
  \label{FIG:Theory}
\end{figure}

The CARS intensity enormously varies between different chiralities due to other resonance conditions. 
%In Figure \ref{FIG:Mtd} we compare the total efficiency of the er-CARS.
Figure \ref{FIG:Theory} shows the dependence of the enhancement at the LO phonon energy as a function of nanotube exciton energy $E_{11}$. The CARS intensity is normalized on laser powers $P_p^2 P_S$ and calibrated by the nanotube concentration, see Methods. The highest signals were provided by the (7,5), (6,5), and (8,3) SWCNTs, with $E_{11}$ located between $E_S$ and $E_p$. The intensity changes between the chiralities were drastic, the (8,3) SWCNTs provide 10$^4$ higher intensity than (10,3). Further, we did not find any non-linear signal from (6,4) SWCNTs, despite clear Raman signatures from the SWCNT in the measured spot. The $E_{11}$ of (6,4) is on the high energy shoulder and lies between $E_p$ and $E_a$. From Figure \ref{FIG:Theory}, we can see that experimental er-CARS intensity is not symmetrically distributed between Stokes, anti-Stokes, and Pump energies.

%The CARS intensity was calibrated on the number of CNTs at a given measurement spot using Raman spectroscopy. Such calibration is justified as G mode Raman matrix elements do not vary much between different chiralities \cite{Jiang2007a}, and the bundles are evenly oriented in all directions. Therefore Raman signal is directly proportional to the CNTs number, and Raman intensity behavior with excitation energy is well known\cite{Duque2011,Gordeev2017,PhysRevB.99.045404}. The other factor we must consider is different scaling to the number of Raman scatters $N$, where Raman scattering scales with $\sim N$ whereas CARS $\sim N^2$. Again we use here the corrected intensity of the G mode. In (6,5), (8,3), and (7,5) enriched samples, the LO mode intensity was comparable, and only in the (10,3) SWCNTs it was much smaller, which creates an additional factor of 42. The calibrated intensities of the CARS at the LO phonon for each nanotube are plotted in Figure \ref{FIG:Theory}. It is possible to observe that the non-linear signal enhancement at the LO phonon is not equally distributed between Stokes, anti-Stokes, and Pump energies.

%Fig4%

Sixth-order perturbation theory yields asymmetric enhancement profile, similar to experimental results. The calculated $I_{FWM}$ is plotted by lines as a function of $E_{11}$ in Figure \ref{FIG:Theory}. The calculation is performed using Eqs. \eqref{EQ:sum} and \eqref{EQ:member} with constant matrix element approximation and damping factors set to $\gamma_{11}=$ 40 meV, typical for the $E_{11}$ excitonic transition \cite{PhysRevB.99.045404}. The values for the lasers energies are $E_S=$1.165 eV, $E_p=$1.36 eV, and for the phonon is $E_{ph}=$0.196 eV are shown by vertical lines. Figure \ref{FIG:Theory} shows that the er-CARS intensity is enhanced when the $E_{11}$ transition energies match the Stokes, pump, or anti-Stokes beam energies. However, the distribution of intensities between these peaks is asymmetric due to the interference effect. The interference is positive at lower energies between $E_S$ and $E_p$, whereas at higher energies between $E_p$ and $E_a$, the interference is negative, and almost no er-CARS enhancement is expected. This nicely reproduces the experimental observations. 
 
%Fig5%
\begin{figure}
  \centering
  \includegraphics[width=16cm]{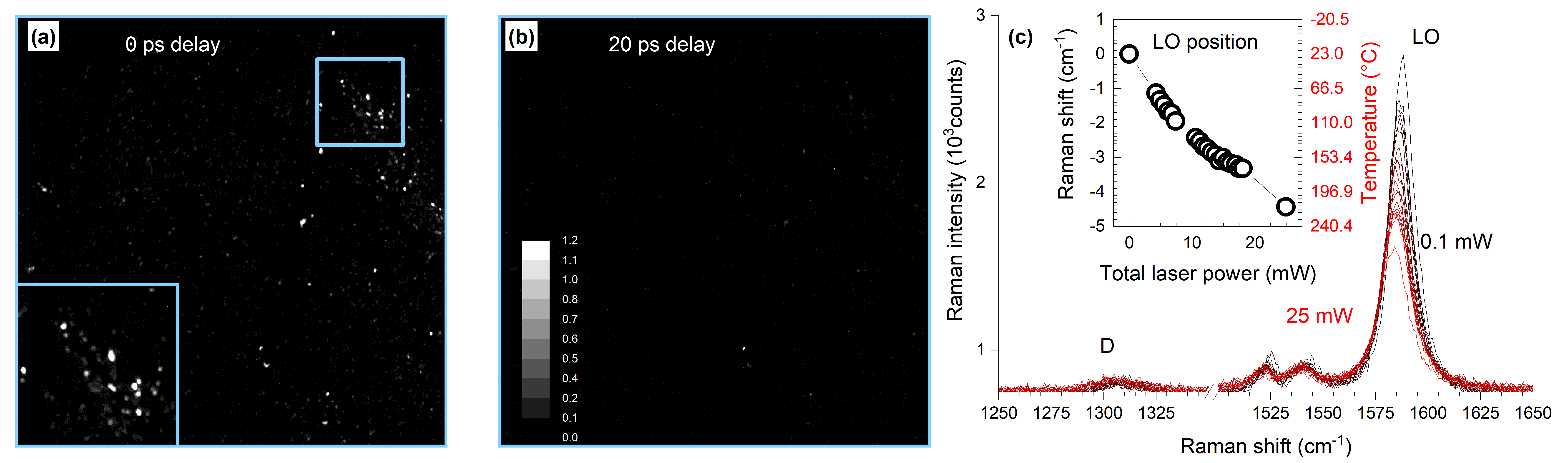}
  \caption{er-CARS microscopy and local heating of the (6,5) SWCNTs. (a) Lateral spatial maps of the er-CARS signal with zero delay, collected within 16 s ($E_S=$1.165 eV, $E_p=$1.36 eV). (b) Same image as (a), but scanned with 20 ps delay. (c) Monitoring local temperature by higher irradiation power $P_{total} = P_{S}+P_{p}$, the Raman excited with 2.2 eV laser, with constant 500\,$\mu W$ power. Inset shows the position of the LO phonon as a function of $P_{total}$; the right scale extracted local temperature using -0.023 cm$^{-1}$/K coefficient. \cite{Ouyang2004}}
  \label{FIG:Appl}
\end{figure}

%\textcolor{red}{I think this table could be in Supplementary Material. But if there is no other material to put in Supp. The material we can keep it in the text.}

We use (6,5) nanotube to perform er-CARS microscopy since most enriching techniques target that species and shows positive er-CARS interference effects. Figure \ref{FIG:Appl}a shows an er-CARS image at the phonon resonance with 200$\times$200 $ \mu m$ in size of (6,5) SWCNTs drop cast onto a quartz substrate. The bright spots correspond to the individual (6,5) SWCNTs or small bundles. The scan is performed within 16 s. In total, 10$^6$ pixels were collected, resulting in the average speed of 16 $\mu s$ per pixel, where the galvanometric mirrors mainly limit that speed. We confirm that the bright spots originate from the er-CARS process. We scanned the same image with a 20 ps delay between the  $E_S$ and $E_p$ pulses, $\sim$ 10 times $\tau_{ph}$. As shown in Figure \ref{FIG:Appl}b, the bright spots disappear, validating the observed signal's er-CARS nature. We further tested the 3D imaging option by tuning the focus outside of the quartz surface, see Supplementary Video 1. Out of focus, the bright spots lose the intensity and convert to circular shapes, enabling the resolution out of plane. Our results show that the er-CARS process enables quick and precise localization of SWCNTs in 2D and 3D, enabling its use for imaging CNT in vivo studies, where Raman or photoluminescence signals from individual species are still relatively weak.

The CNTs can be used for the simultaneous er-CARS imaging  and local heating of the CNT for targeted destruction of viruses and cancer cells.\cite{Golubewa2020} Most biological tissues have a transparency window in the IR region where the $E_{11}$ transition energies of small-diameter CNTs are found. Therefore the heat wave can be localized near the CNT position. Now the temperature control methods rely primarily on mathematical models.\cite{Eskiizmir2017,Golubewa2020} In our setup, we can directly probe and control the local temperature using the Raman signal of CNT as a thermometer. We measured the Raman spectrum of a small CNT bundle during simultaneous irradiation with pulsed $E_p$ and $E_S$ beams of varying powers. The Raman spectra are shown in Figure \ref{FIG:Appl}c from weak (black) to strong (red) irradiation powers (the measured laser power is the sum of both pump and Stokes beams). Notably, no defects were introduced to the sample, as the D mode doesn't grow in intensity\cite{Gordeev2016a}.
Further, we observe that the LO phonon broadens and shifts to smaller energies. We plot the LO phonon position as a function of total power $P_S+P_p$ in the inset of Figure \ref{FIG:Appl}c. The Raman shift is converted to temperature by a -0.023 cm$^{-1}$/K coefficient \cite{Ouyang2004}. We can quickly achieve temperatures close to 200 °C when even 80 °C will denature the proteins in the living cells\cite{Chen2016}. This means that CNT-based er-CARS microscopy is suitable for tracking, local heating, and disrupting cancer cells.

\section{Conclusions} 
We demonstrated that phonon and excitonic resonances can enhance four-wave mixing signals by a few orders of magnitude in SWCNTs, giving rise to the er-CARS process. We investigated the resonant signals in five different CNT chiralities. The intensity varied drastically and depended upon the relative energy between the pump and Stokes beams concerning the first excitonic transition. We developed a theory based on quantum interference between 43 scattering pathways, which delivered a qualitative agreement with experimental results. Further, we tested the applicability of resonant wave mixing and imaged individual (6,5) CNTs over 200x200 $\mu m^2$ quartz substrate in a time of 16 s. The time was only limited by the speed of our galvanometric mirrors, and even quicker scans could be achieved in the future. As a possible application, the simultaneous er-CARS imaging with local heating with lasers used in CARS could be used for targeting and cell disruption in the future. Our work demonstrates that single-walled carbon nanotubes are exciting materials for resonant non-linear light-matter interaction with potential applications in tracking and nanotoxicology.

\section{Acknowledgments}
\begin{acknowledgments}
G.G. gratefully acknowledges the German Research Foundation (DFG via SFB 658, subproject A6). Dahlem research school (FU bright fellowship) and Focus Area NanoScale of Freie Universitaet Berlin. L.L. gratefully acknowledges the Alexander von Humboldt Foundation for its financial support. B.F. acknowledges DFG grants FL 834/5-1, FL 834/7-1, FL 834/9-1 and FL 834/12-1. L.L, A.J., and L.M.M primarily edge financial support from CNPq, CAPES, FAPEMIG, FINEP, Brazilian Institute of Science and Technology (INCT) in Carbon Nanomaterials and Rede Mineira de Materiais 2D (FAPEMIG). L.M.M. also acknowledges the CAPES and Humboldt fellowship. All authors are thankful for the fruitful discussions with Prof. Riichiro Saito and Prof. Stephanie Reich and acknowledge help with language corrections done by Oisín Garrity. 

\end{acknowledgments}

\bibliography{er_CARS.bib}

\end{document}